\documentclass[twocolumn,showpacs,preprintnumbers,amsmath,amssymb]{revtex4}

\usepackage{graphicx}
\usepackage{dcolumn}
\usepackage{bm}
\usepackage{subfigure}

\begin{document}

\title{Magnetic field induced semimetal-to-canted-antiferromagnet transition\\on the honeycomb lattice}

\author{M.~Bercx, T.~C.~Lang and F.~F.~Assaad}
\affiliation{Institut f\"ur Theoretische Physik und Astrophysik, Universit\"at W\"urzburg, Am Hubland, D-97074 W\"urzburg, Germany}

\date{\today}
\begin{abstract}
It is shown that the semimetallic state of the two-dimensional honeycomb lattice with a point-like Fermi
surface is unstable towards a canted antiferromagnetic insulator upon application of an in-plane
magnetic field. This instability is already present at the mean-field level; the magnetic field shifts the up- and the down-spin cones in opposite directions thereby generating a finite density of states at the
Fermi surface and a perfect nesting between the up- and the down-spin Fermi sheets. This perfect nesting
triggers a canted antiferromagnetic insulating state. Our conclusions, based on mean-field arguments,
are confirmed by auxiliary field projective quantum Monte Carlo methods on lattices up to $12 \times
12$ unit cells.
\end{abstract}

\maketitle
\section{\label{sec:section1}Introduction}
Graphene, or the physics of electrons on the honeycomb lattice,  has recently received tremendous
attention due to its  semimetallic nature with low-energy quasiparticles behaving as massless Dirac
spinors; a recent review can be found in Ref. \cite{CastroNeto09}. A crucial point is the stability of
this semimetallic  phase to  particle-hole pairing. In particular, research activities have been devoted
to the investigation of  magnetic field induced transitions as a function of magnetic fields
\cite{Aleiner07,Kempa03,Kharzeev06} and electronic correlations.
\cite{Sorella92,Paiva05,Herbut06,Raghu08}     The vanishing density of states at the  Fermi energy
protects the semimetallic state against   {\it weak} correlations. Notably, a  finite critical value
of the repulsive Hubbard interaction $U/t$ is required to destabilize the  semimetallic state in favor of an antiferromagnetic Mott insulator. \cite{Sorella92,Paiva05,Herbut06}

In this paper, we will argue that the semimetallic state is unstable against the application of an
in-plane magnetic field.  The mechanism behind this  instability can be understood already at the
mean-field level. \cite{Milat04, Beach04, Peres04} The magnetic field generates a finite Fermi surface density of states. A Stoner
instability for arbitrarily small values of the  Coulomb repulsion arises from  the perfect nesting of
the  spin split Fermi sheets. This triggers antiferromagnetic order  with staggered magnetization
perpendicular to the  applied magnetic field  and the opening of a charge gap. That the application of
an in-plane magnetic field in the continuum limit facilitates a spontaneous symmetry breaking  has
already been pointed out in Ref. \cite{Kharzeev06}.  The purpose of this paper is to show that those
mean-field arguments indeed capture the correct physics, since exact  quantum Monte Carlo (QMC) simulations on
the honeycomb lattice compare favorably with those mean-field results.

The outline of the paper is as follows. In Sec.~\ref{sec:section2} the model Hamiltonian is introduced
and its mean-field solution is discussed. The projective QMC method for ground
state properties and numerical results are presented in Sec.~\ref{sec:section3}. The last section,
Sec.~\ref{sec:section4}, contains the summary and the conclusions, also with respect to the
experimental relevance of our findings.
\section{\label{sec:section2}Model Hamiltonian and mean-field treatment}

\begin{figure}[htbp]
   \centering
   \includegraphics[width=0.40\textwidth]{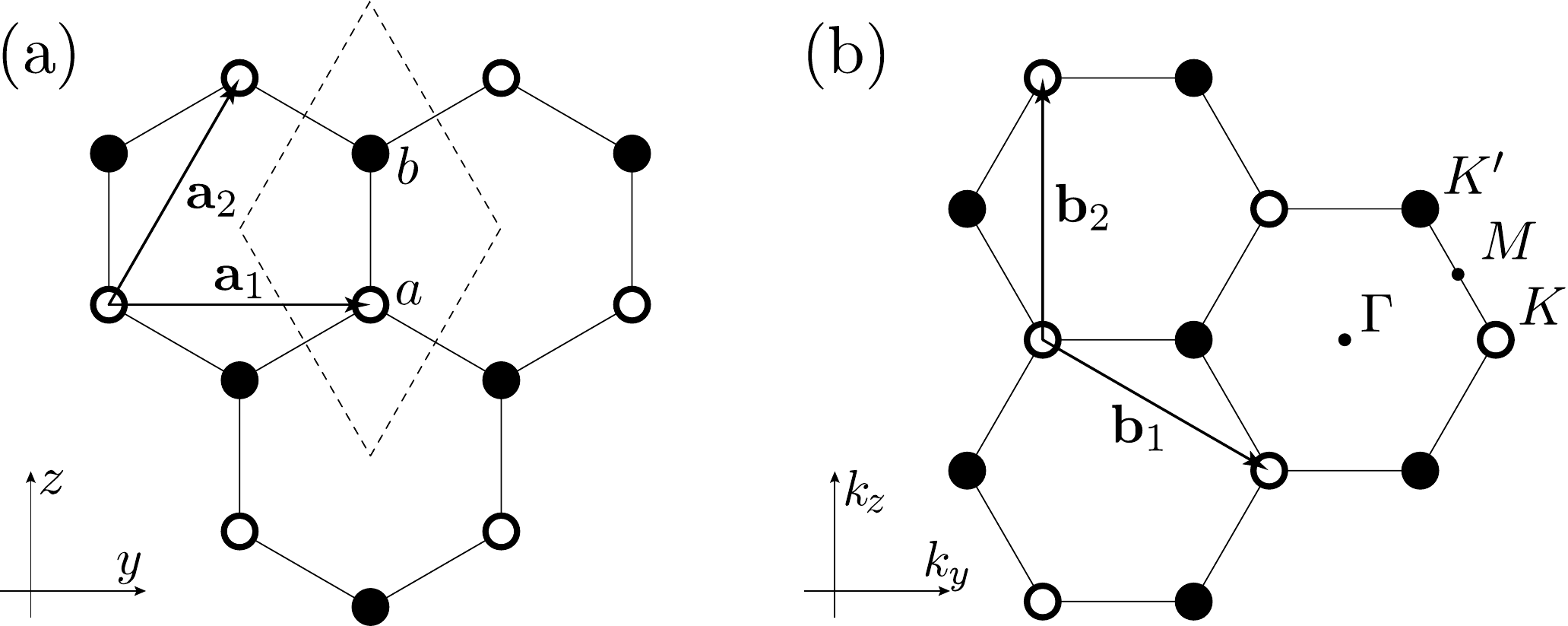}
   \caption{(a) Real and (b) reciprocal space lattice vectors of the honeycomb lattice, 
   ${\mathbf{a}_{1}=a_0\;(0,1,0)}$, ${\mathbf{a}_{2}=a_0\;(0,1/2,\sqrt{3}/2)}$ and
   ${\mathbf{b}_{1}=2\pi/a_0\;(0,1,-1/\sqrt{3})}$, ${\mathbf{b}_{2}=2\pi/a_0\;(0,0,2/\sqrt{3})}$, 
   with $a_0$ being the lattice constant. The unit cell with the orbitals $a$ and $b$ is indicated
   by the dashed diamond shape. Filled (empty) circles denote sites on the same sublattice.
   \label{fig1}}
\end{figure}

Our starting point is the Hubbard model on the  honeycomb lattice shown in  Fig.~\ref{fig1}.
\begin{eqnarray}\label{eqnham}
H&=&H_{0}+H_{U}+H_{B},\nonumber\\
H_0&=& -t\sum_{{\bf i},{\bf r}, \sigma}\Big( \hat{a}^{\dagger}_{{\bf i},\sigma}\hat{b}_{{\bf i} + {\bf r},
   \sigma}  +  \hat{b}^{\dagger}_{{\bf i} + {\bf r},\sigma}  \hat{a}_{{\bf i},\sigma} \Big), \nonumber  \\
H_U&=&  U \sum_{l=a,b}\sum_{{\bf i}} \left( \hat{n}_{{\mathbf{i},l},\uparrow}   - 1/2 \right)
                                 \left( \hat{n}_{{\mathbf{i},l},\downarrow} - 1/2 \right),  \nonumber\\
H_B&=& \frac{g}{2}\mu_{\mathrm{B}}B\sum_{l=a,b} \sum_{{\bf i},\sigma}p_{\sigma}\hat{n}_{{\mathbf{i},l},\sigma}\;.
\end{eqnarray}
The electron operator $\hat{a}^{\dagger}_{{\bf i},\sigma}$ ($\hat{b}^{\dagger}_{{\bf i},\sigma}$)
creates an electron on the orbital $a$ ($b$) in the unit cell ${\bf i} $ and the associated
electron-density  operator is
${\hat{n}^{l}_{{\bf i},\sigma}=\hat{a}^{\dagger}_{{\bf i},\sigma}\hat{a}_{{\bf i},\sigma}}$
(${\hat{b}^{\dagger}_{{\bf i} ,\sigma}\hat{b}_{ {\bf i},\sigma}}$), for $l=a$ ($b$).
Owing to the bipartite nature of the lattice, hopping with matrix element $t$ occurs only between the $a$ and the $b$ orbitals of unit cells related by lattice vector ${\bf r }$ in the three directions $\left\{ {\bf 0}, {\bf a}_1 - \bf{a}_2,  -{\bf a}_2 \right\}$.
The on-site electron-electron repulsion is denoted by  $U>0$ and $p_{\sigma}=\pm 1 $ for
$\sigma=\uparrow,\downarrow$. In the present case of
half filling the chemical potential vanishes. In the following, we set $(g/2)\mu_{\mathrm{B}}\equiv 1$.
We have included only a Zeeman coupling to the magnetic field.  Hence, for comparison with
experiments, we can only consider setups with magnetic field orientations
parallel to the lattice plane since only in this case we can neglect the orbital coupling.

\begin{figure}[htbp]
   \centering
   \includegraphics[width=0.40\textwidth]{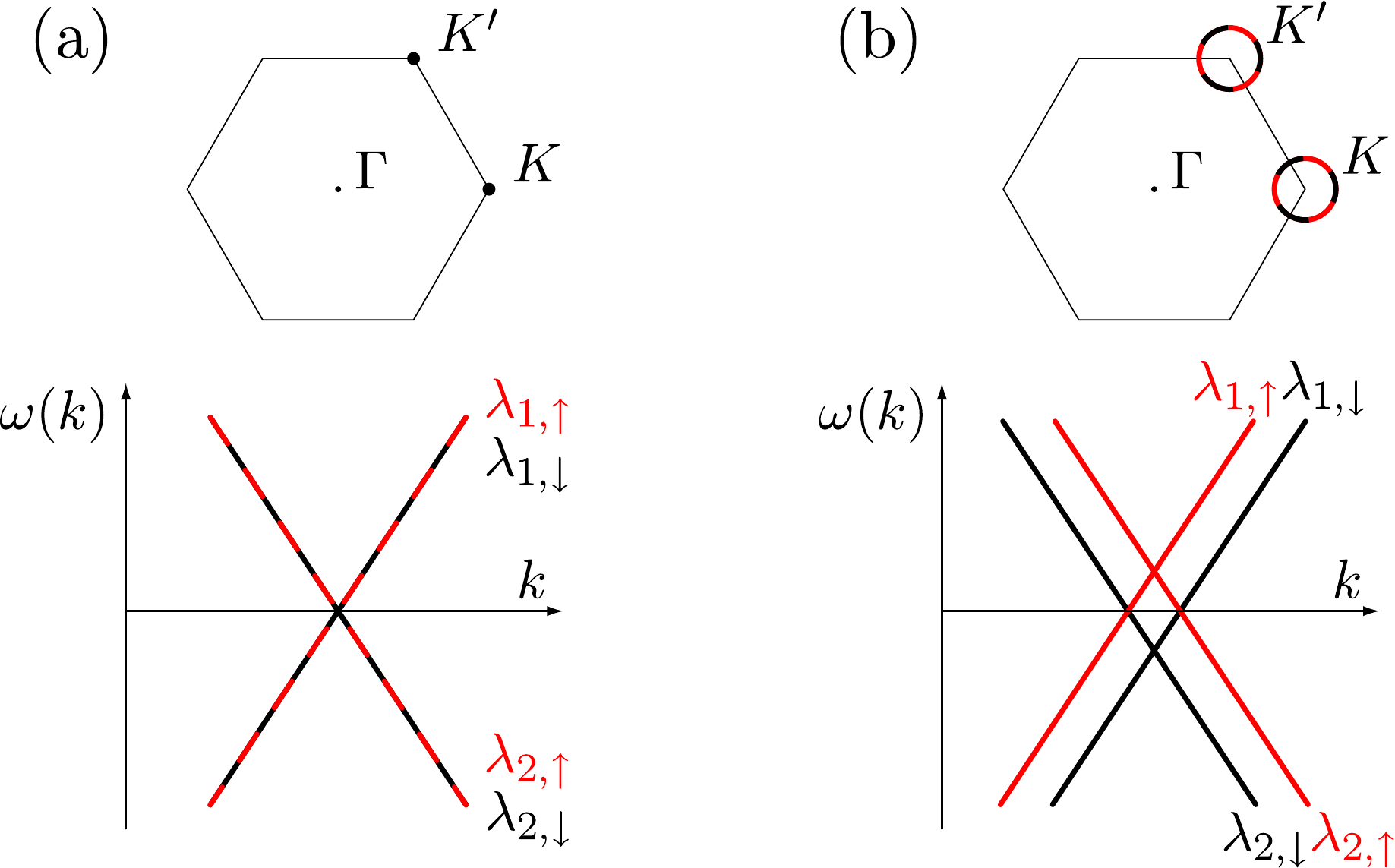}
   \caption{ Visualization of the nesting of spin-up and spin-down Fermi surfaces. In case of (a) $B=0$
   the spin bands collapse onto each other, whereas for (b) $B>0$ the bands are shifted by virtue of
   the magnetic field leading to the nesting relation ${\lambda_{1,\uparrow}(\mathbf{k})=
   - \lambda_{2,\downarrow}(\mathbf{k})}$.
   \label{fig2}}
\end{figure}

The Hamiltonian $H_0$ gives rise to two bands,  
\begin{equation}
     \lambda_n( {\bf k} ) =  p_n  \left| t  \sum_{\bf r} e^{-i {\bf k} \cdot {\bf r} } \right|,
\end{equation}
with  $p_n = \pm 1$  for $n=1,2$, respectively. At half-band filling the Fermi surface
consists of two  points, $K,K'$ in Fig.~\ref{fig2},   with density of states
\begin{equation}
     \rho(\omega) = \frac{1}{N} \sum_{\bf k} \delta \left( \omega - \lambda_{1}[{\bf k}] \right).
\end{equation}
Here, $N$ corresponds to the lattice size and  linearization of  the dispersion relation
around $K$ and $K'$ yields: $ \rho(\omega) \propto \omega/t^2 $ for $ \omega\ll t $.
\begin{figure}[tbp]
   \centering
   \includegraphics[width=0.45\textwidth]{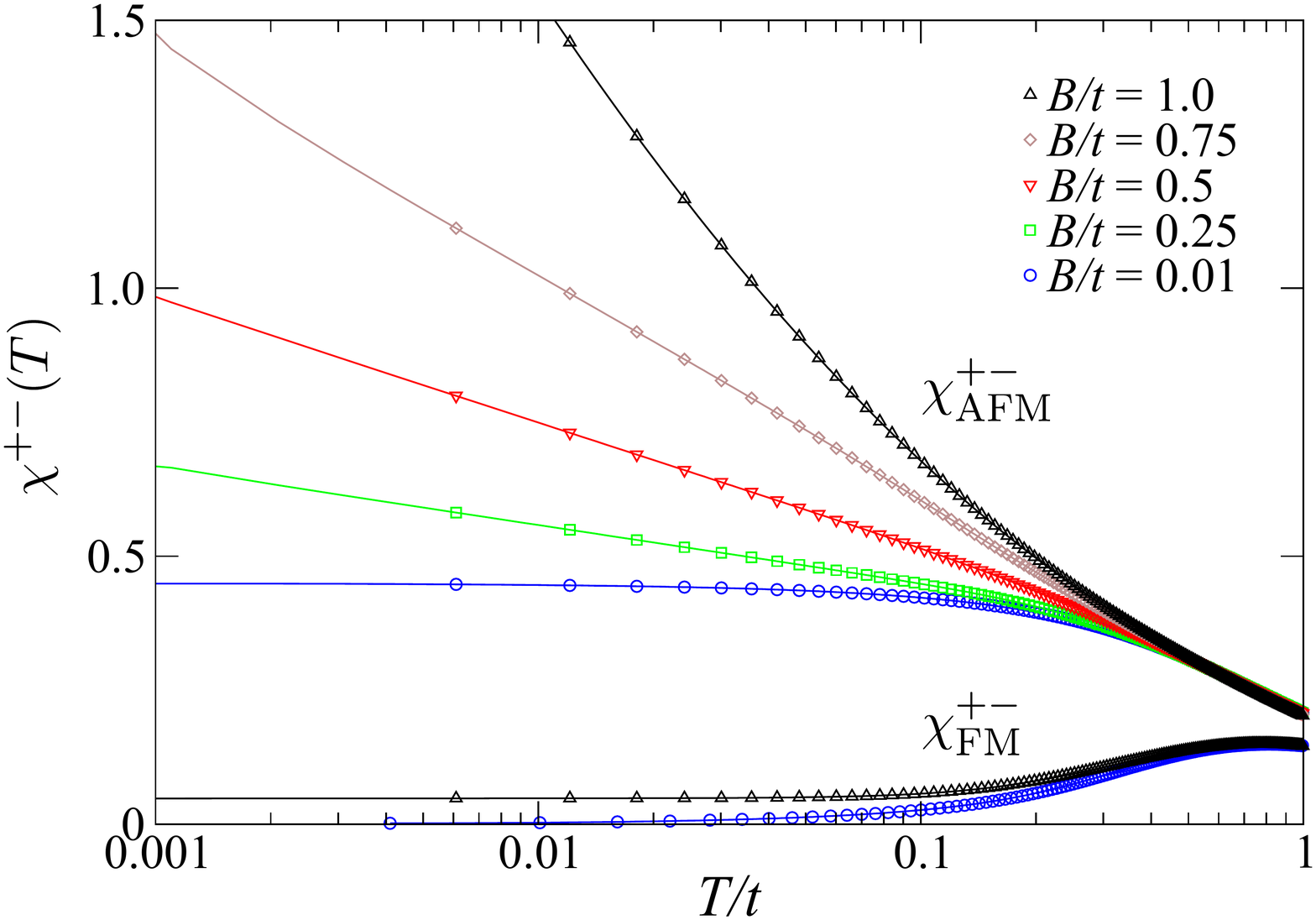}
   \caption{(Anti-)ferromagnetic susceptibilities ${\chi^{+-}_{\mathrm{(A)FM}}(\mathbf{Q})}$ for the magnetic fields $B/t=0.01,...,1.0$.
   \label{fig3}}
\end{figure}

Prior to examining the mean-field Hamiltonian we demonstrate the instability of the non-interacting
system when turning on the magnetic field.  Consider the transverse magnetic susceptibility tensor:
\begin{equation}\label{eqnchi}
\begin{pmatrix}
\chi^{+-}_{aa}(\mathbf{q}) & \chi^{+-}_{ab}(\mathbf{q})\\ \chi^{+-}_{ba}(\mathbf{q}) & \chi^{+-}_{bb}(\mathbf{q})
\end{pmatrix}=
\begin{pmatrix}
\chi^{+-}_{d}(\mathbf{q}) & \chi^{+-}_{o}(\mathbf{q})\\ \chi^{+-}_{o}(\mathbf{q}) & \chi^{+-}_{d}(\mathbf{q})
\end{pmatrix}\;.
\end{equation}
Here 
$\chi^{+-}_{l,l'}(\mathbf{q})=\int_{0}^{\beta}d\tau \langle S^{+}_{l}(\mathbf{q},\tau)
S^{-}_{l'} (-\mathbf{q},0) \rangle $ and the indices $l,l' = a,b$ label the sublattices. $\beta$ corresponds to the inverse temperature  and the
spin raising  and lowering operators read  $S^{+}_{a} (\mathbf{q}) =  \frac{1}{\sqrt{N}}
\sum_{{\bf i} } e^{-i {\bf q} \cdot {\bf i} } \hat{a}^{\dagger}_{{\bf i},\uparrow}
\hat{a}_{{\bf i},\downarrow} $
and  
$S^{-}_{a} (\mathbf{q}) =  \frac{1}{\sqrt{N}}
\sum_{{\bf i} } e^{-i {\bf q} \cdot {\bf i} } \hat{a}^{\dagger}_{{\bf i},\downarrow}
\hat{a}_{{\bf i},\uparrow} $,  respectively.  Similar definitions hold for the $b$ sublattice.  
The  eigenvectors of the magnetic susceptibility tensor are given by
\begin{equation}
\begin{gathered}    
       \chi^{+-}_{\mathrm{FM}}(\mathbf{q})  = \chi^{+-}_{d}(\mathbf{q}) +  \chi^{+-}_{o}(\mathbf{q}),    \\
       \chi^{+-}_{\mathrm{AFM}}(\mathbf{q})  = \chi^{+-}_{d}(\mathbf{q}) -  \chi^{+-}_{o}(\mathbf{q}),  
\end{gathered}    
\end{equation}
and correspond, respectively, to the ferromagnetic and the antiferromagnetic alignments of spins
within the  unit cell.  Nesting  occurs at  $\mathbf{q} =  \mathbf{Q} = (0,0) $  and leads to a
logarithmic  divergence of the antiferromagnetic  mode  at finite values of the magnetic field (Fig.~\ref{fig3}).
\begin{subequations}
\label{bothsus}
\begin{eqnarray}
   \label{fmsus}
   \chi^{+-}_{\mathrm{FM}}\left( {\mathbf{Q}} \right) & = &
      \frac{1}{4NB}\sum_{\mathbf{k}}\sum_{n=1}^{2}\Big(f\big(\lambda_{n,\downarrow}(\mathbf{k})\big)-f\big(\lambda_{n,\uparrow}(\mathbf{k})\big)\Big)\nonumber\\
      & = &\frac{1}{4B}\int d\omega\Big[\rho(\omega-B)-\rho(\omega+B)\Big]\nonumber\\
      &   &\times\;\Big(f(-\omega)-f(\omega)\Big)\;,
\end{eqnarray}
and
\begin{eqnarray}
   \label{afmsus}
   \chi^{+-}_{\mathrm{AFM}}\left( {\mathbf{Q}} \right) & = &
   \frac{1}{2N}\sum_{\mathbf{k}}\sum_{\substack{n,m=1\\ n\neq m}}^{2}
   \frac{f\big(\lambda_{n,\downarrow}(\mathbf{k})\big)-f\big(\lambda_{m,\uparrow}(\mathbf{k})\big)}{\lambda_{m,\uparrow}(\mathbf{k})-\lambda_{n,\downarrow}(\mathbf{k})}\nonumber\\
   & = &\frac{1}{4}\int d\omega \Big[\rho(\omega-B)+\rho(\omega+B)\Big]\nonumber\\
   &   &\times\;\frac{f(-\omega)-f(\omega)}{\omega}\;.
\end{eqnarray}
\end{subequations}
In the above,
\begin{equation}
     \lambda_{n,\sigma} ({\bf k}) = \lambda_{n} ({\bf k})   + p_{\sigma} B
\end{equation}
are single-particle states of $H_0 + H_B$
and $f(\omega) = \frac{1}{1 + e^{\beta \omega} } $ is the Fermi function.
In the low-temperature limit, one can approximate  the integral of Eq.~(\ref{afmsus}) to
obtain
\begin{equation}\label{afmsus2}
\chi^{+-}_{\mathrm{AFM}}({\bf Q})  \propto
\begin{cases}
\rho(B)\ln\Big(\frac{W}{2 k_{\mathrm{B}}T}\Big), &|B|>0 \\
\text{constant}, &B=0\;,
\end{cases}
\end{equation}
where $W$ corresponds to the bandwidth.
Clearly, the  divergence of the transverse  susceptibility  in the antiferromagnetic channel
stems from the  nesting property, ${\lambda_{1,\uparrow}(\mathbf{k})=
- \lambda_{2,\downarrow}(\mathbf{k})} $. At zero
magnetic field, this instability is  cut off by the vanishing density of sates  $\rho(\omega) \propto
\omega/t^2$.  At $B > 0$ the low-energy density of states is finite thereby revealing the
nesting instability.

Given the above instability, the mean-field Hamiltonian is derived by assuming the magnetization
${\bf m}$ to be alternating on
the sublattices: $\mathbf{m}_{l}=\begin{pmatrix} 0 , & m_{\perp}(-1)^{l}, & m_{\parallel} \end{pmatrix}$
with the index $l=0,1$ referring to the orbitals in the unit cell. That is, the magnetization
${\bf  m}$  has a constant component $m_{\parallel}$ parallel to the field axis
and a staggered component $m_{\perp}$ in the plane perpendicular to the field.

\begin{figure}[tbp]
   \centering
   \subfigure[]
      {\includegraphics[width=0.45\textwidth]{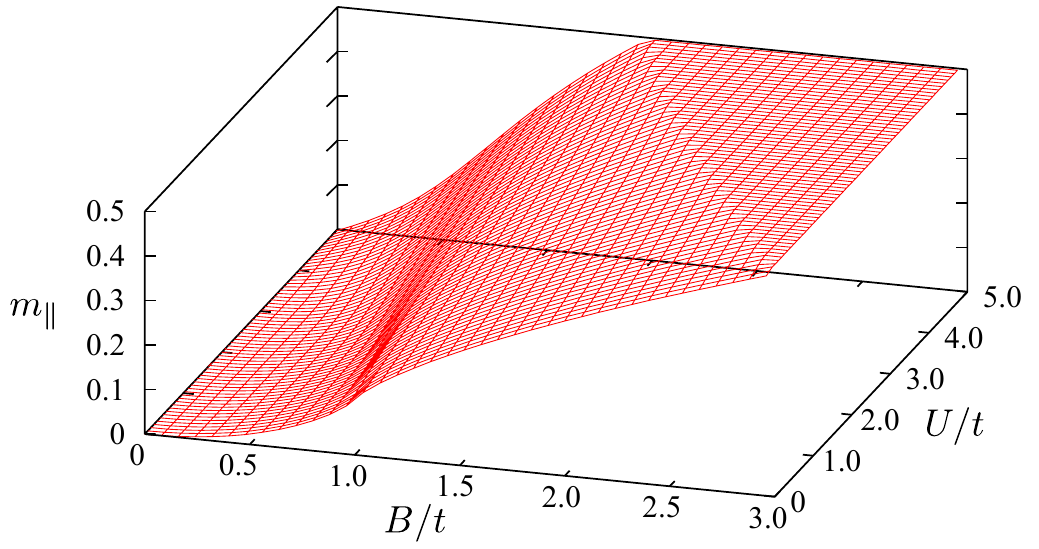}}
   \subfigure[]
      {\includegraphics[width=0.45\textwidth]{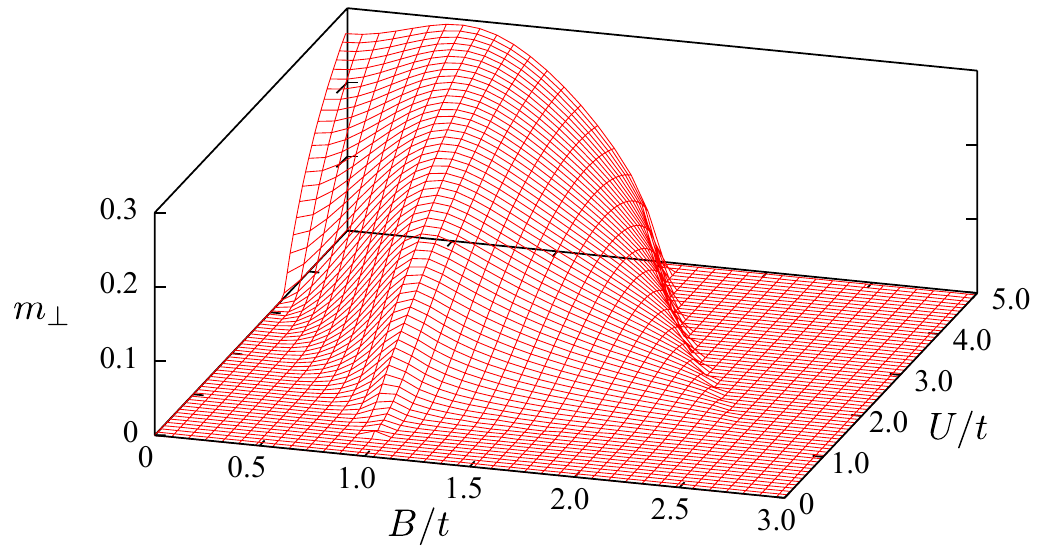}}
   \caption{(a) Parallel magnetization $m_{\parallel}$ and (b) staggered magnetization $m_{\perp}$ vs $U$ and $B$, obtained by numerically solving the gap equations, (\ref{gapeqn}).
   \label{fig4}}
\end{figure}

\begin{figure*}[htbp]
   \centering
   \subfigure[$$]{\includegraphics[width=0.23\textwidth, angle = -90]{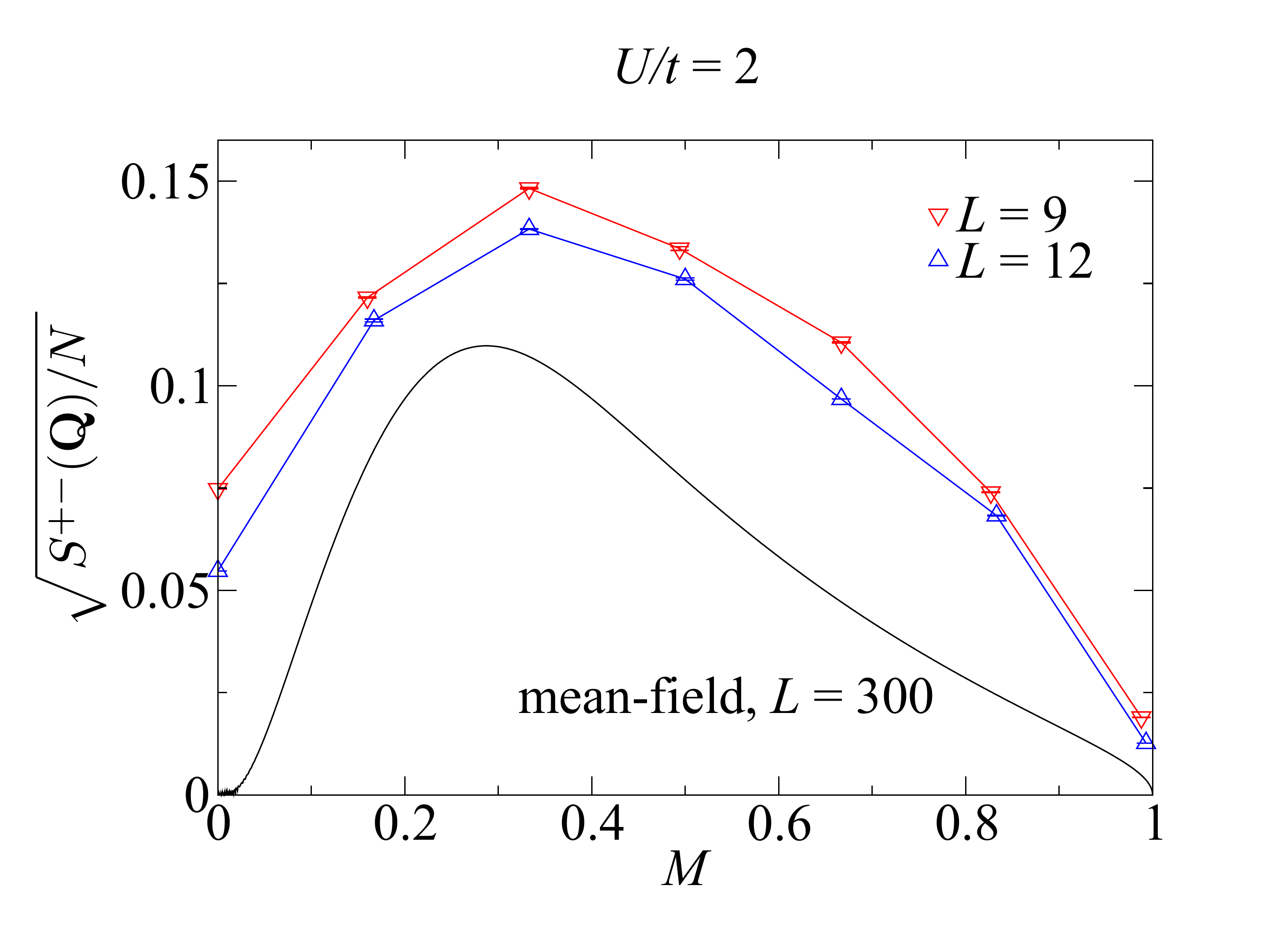}}
   \subfigure[$$]{\includegraphics[width=0.23\textwidth, angle = -90]{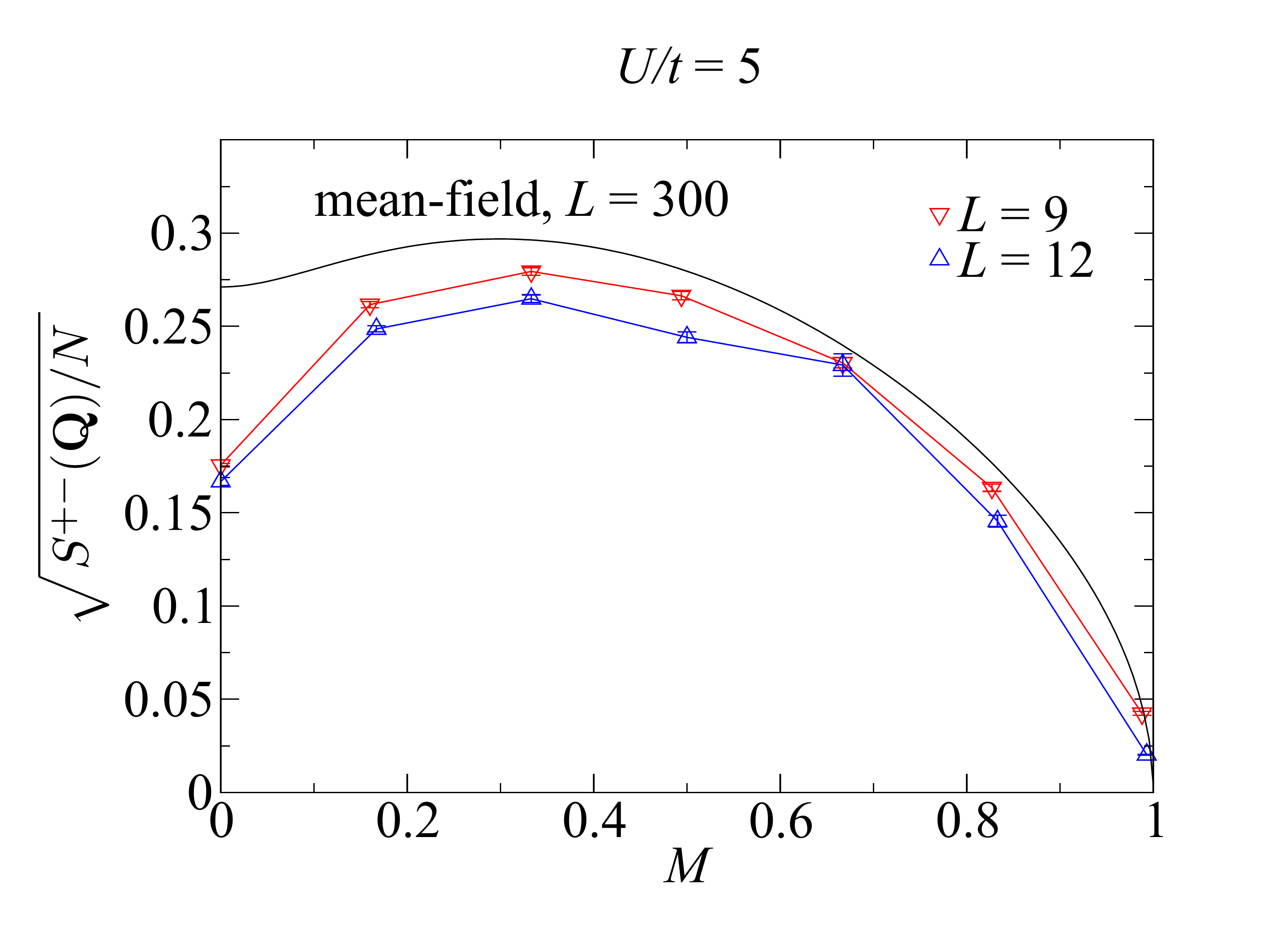}}
   \subfigure[$$]{\includegraphics[width=0.23\textwidth, angle = -90]{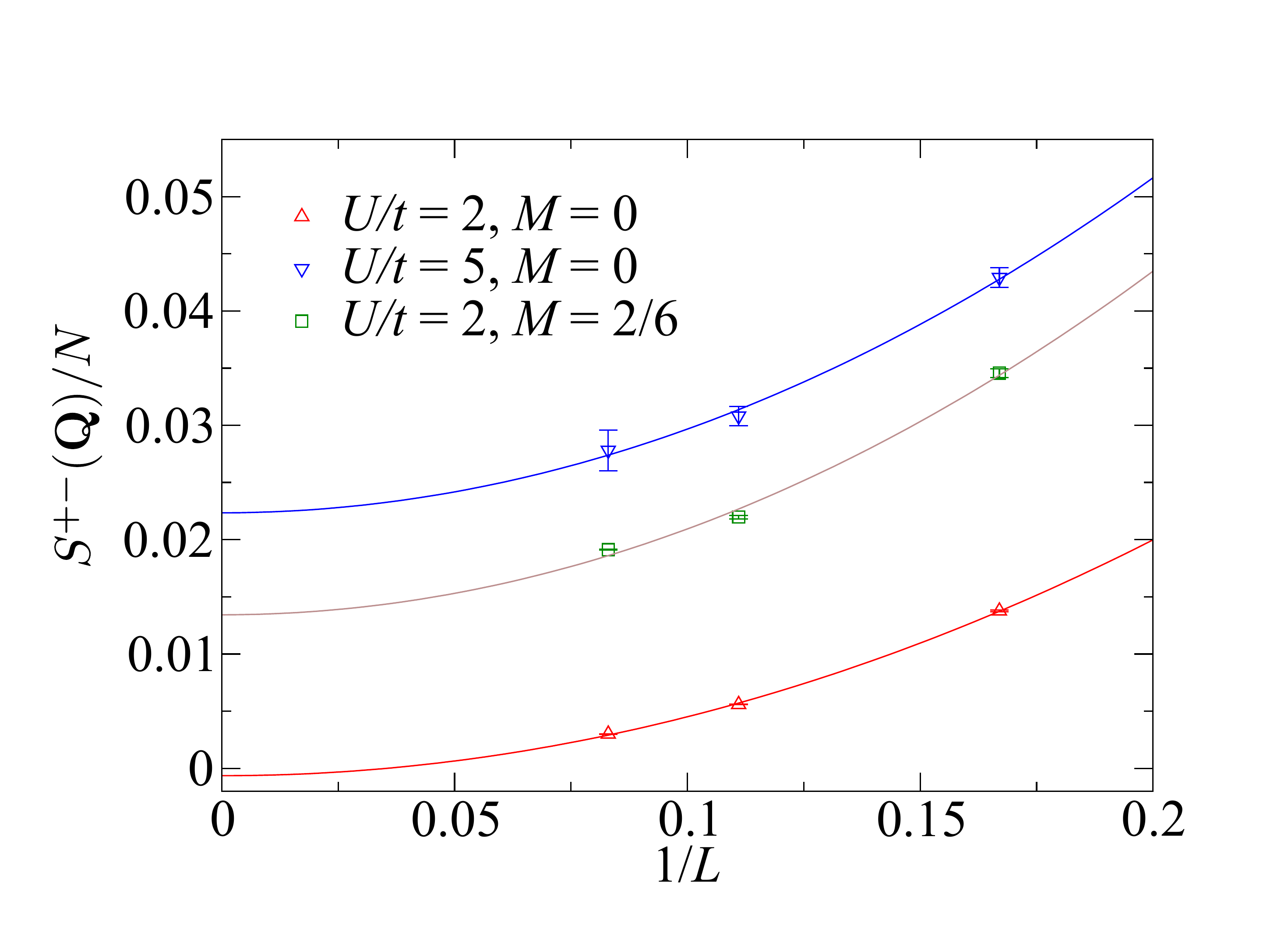}}
   \caption{Staggered magnetization $\sqrt{ S^{+-}(\mathbf{Q})/N }$ (a) below and (b) above the critical interaction strength. (c) Finite-size extrapolation of $S^{+-}(\mathbf{Q})/N$.
In the semimetallic phase,  $U/t = 2$, the data are consitent with the onset  of transverse
staggered order at  finite magnetization $M$. For comparison, we have plotted the $U/t =5  $ data in the absence of a magnetic field.
This value of the Hubbard interaction places us in the antiferromagnetic Mott insulating state.
   \label{fig5}}
\end{figure*}

To decouple the interaction part of the Hamiltonian we rewrite it as ${H_{U} = -\frac{2}{3}U \sum_{l=a,b}
 \sum_{{\bf i}}\hat{ {\bf S} }_{{\bf i},l}^{2}}$, with $\hat{{\bf S}}_{{\bf i},l} $  being
the spin-1/2 operator,  and the fluctuation term of
${\hat{{\bf S}}_{{\bf i},l}^{2}=[{\bf m}_{l}+(\hat{\bf S}_{{\bf i},l}-{\bf m}_{l})]^{2}}$ is omitted.  
With this ansatz, the mean-field  Hamiltonian  mixes the up- and the down-spin sectors thereby yielding
four quasiparticle bands:
\begin{eqnarray}\label{hammf}
H_{\mathrm{MF}} & = &\sum_{\mathbf{k}}{\sum_{n,m=1}^{2}{E_{n,m}(\mathbf{k})\;\hat{\gamma}_{n,m,\mathbf{k}}^{\dagger}\hat{\gamma}_{n,m,\mathbf{k}}}}\nonumber\\
&  & +\;\frac{4}{3}\, U N \big(m_{\parallel}^{2}+m_{\perp}^{2}\big),
\end{eqnarray}
with
\begin{equation}
E_{n,m}(\mathbf{k}) =
p_m \sqrt{\big(\lambda_{n}({\bf k})  + B_{\mathrm{eff}}\big)^{2}+\Delta_{\perp}^{2}}\;.
\end{equation}
Here, ${p_m = \pm 1}$, ${B_{\mathrm{eff}}=B-\Delta_{\parallel}}$,  ${\Delta_{\parallel}=\frac{2}{3}U m_{\parallel}}$, and ${\Delta_{\perp}=\frac{2}{3}U m_{\perp}}$.
Minimizing the free energy with respect to $m_{\parallel}$ and $m_{\perp}$ yields the gap equations
\begin{subequations}\label{gapeqn}
\begin{eqnarray}
1=&&\frac{U}{6N}\sum_{\mathbf{k}}\sum_{n=1}^{2}\frac{1}{\sqrt{\left (\lambda_n(\mathbf{k})+B_{\mathrm{eff}}\right)^{2}+\Delta_{\perp}^{2}}}\nonumber\\
=&&\frac{U}{6}\int d\omega\frac{\rho(\omega+B_{\mathrm{eff}})+\rho(\omega-B_{\mathrm{eff}})}{\sqrt{\omega^{2}+\Delta_{\perp}^{2}}}\label{gapeqn1},\\
m_{\parallel}
=&&\frac{1}{4N}\sum_{\mathbf{k}}\sum_{n=1}^{2}\frac{-\left(\lambda_n(\mathbf{k})+B_{\mathrm{eff}}\right)}{\sqrt{\left(\lambda_n(\mathbf{k})+B_{\mathrm{eff}}\right)^{2}+\Delta_{\perp}^{2}}}\nonumber\\
=&&\frac{1}{4}\int d\omega\frac{\Big(\rho(\omega+B_{\mathrm{eff}})-\rho(\omega-B_{\mathrm{eff}})\Big)\omega}{\sqrt{\omega^{2}+\Delta_{\perp}^{2}}}\label{gapeqn2}.
\end{eqnarray}
\end{subequations}
\begin{figure*}[htbp]
   \centering
   \subfigure[$$]{\includegraphics[width=0.49\textwidth]{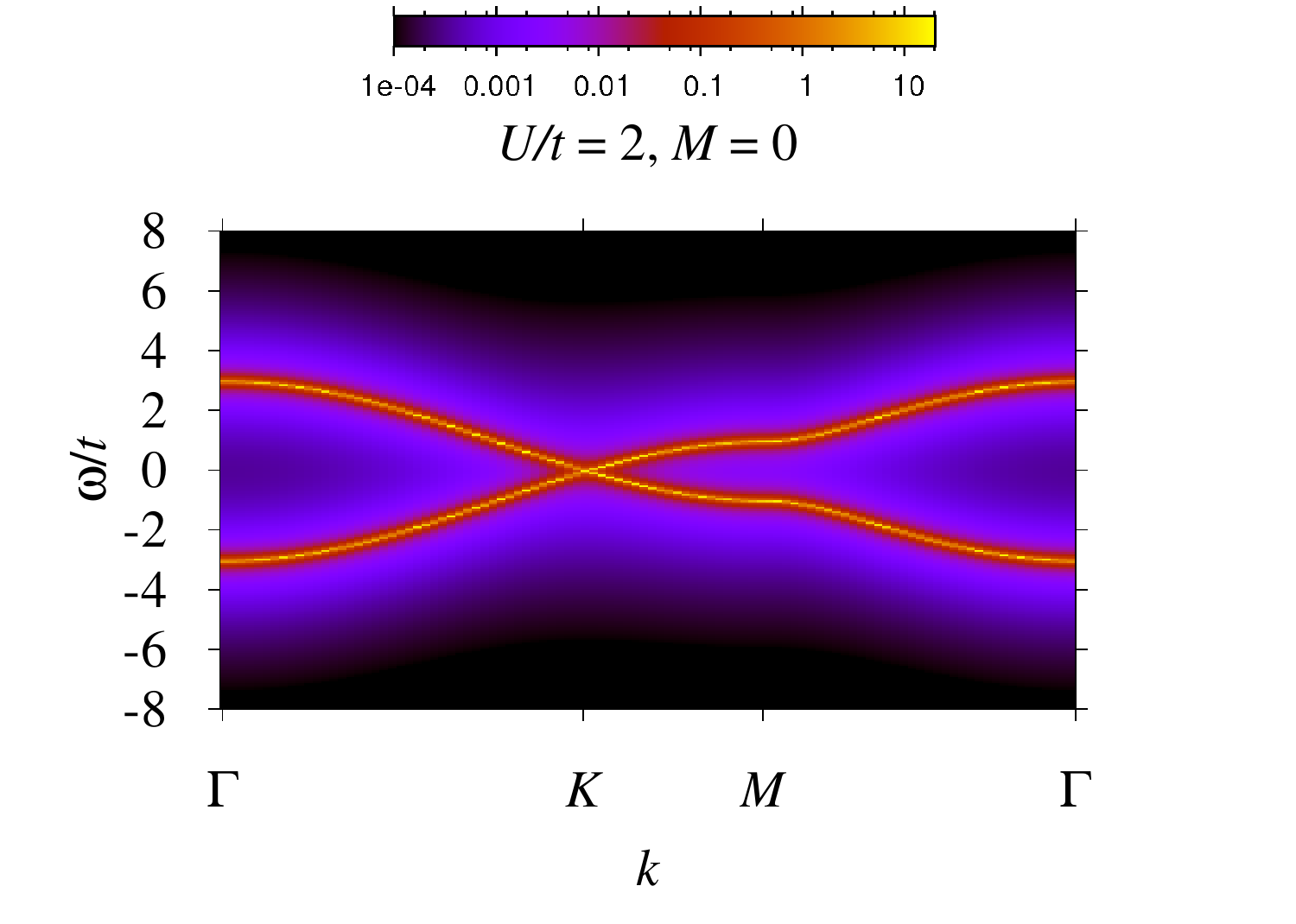}}
   \subfigure[$$]{\includegraphics[width=0.49\textwidth]{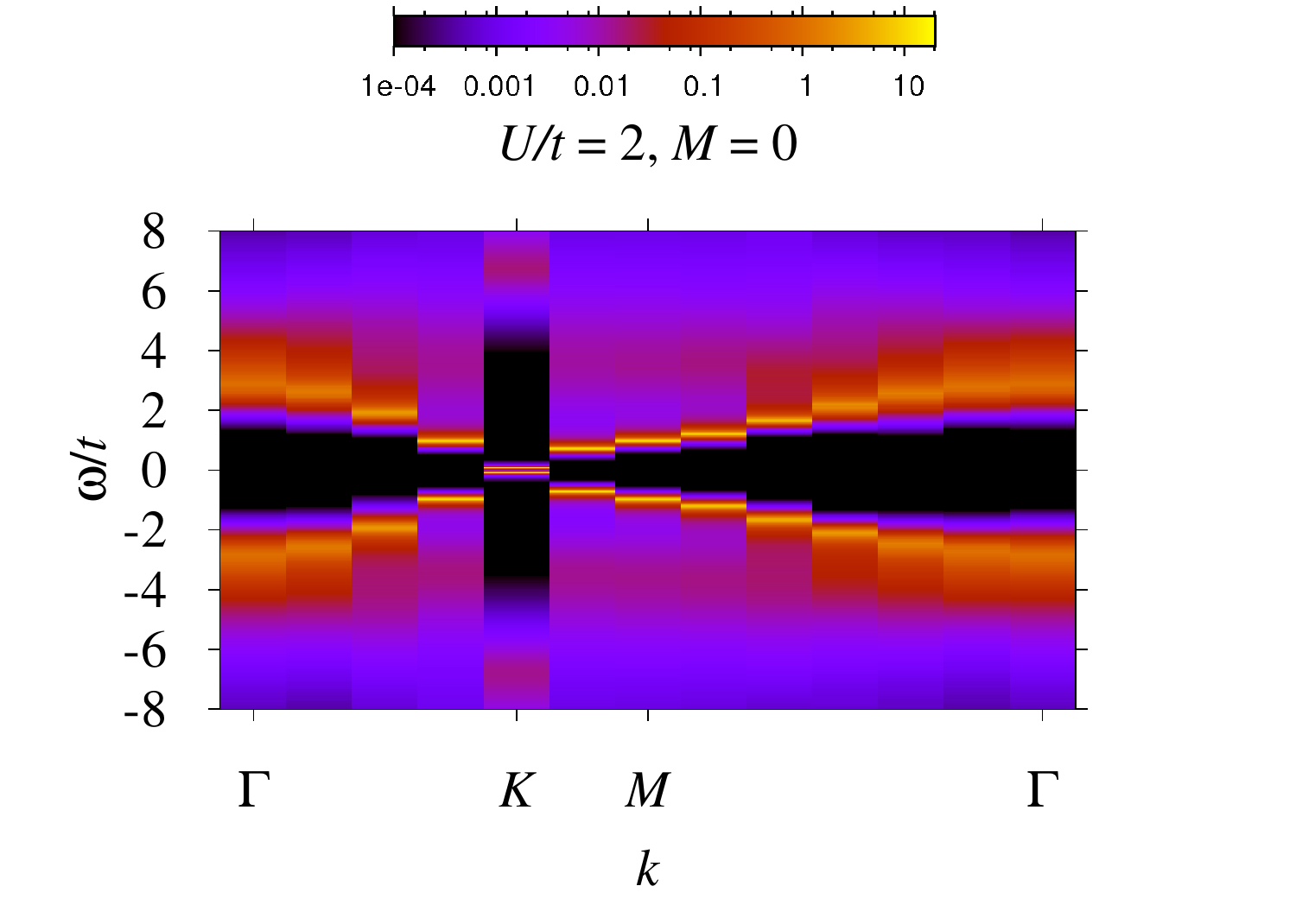}}
   \subfigure[$$]{\includegraphics[width=0.49\textwidth]{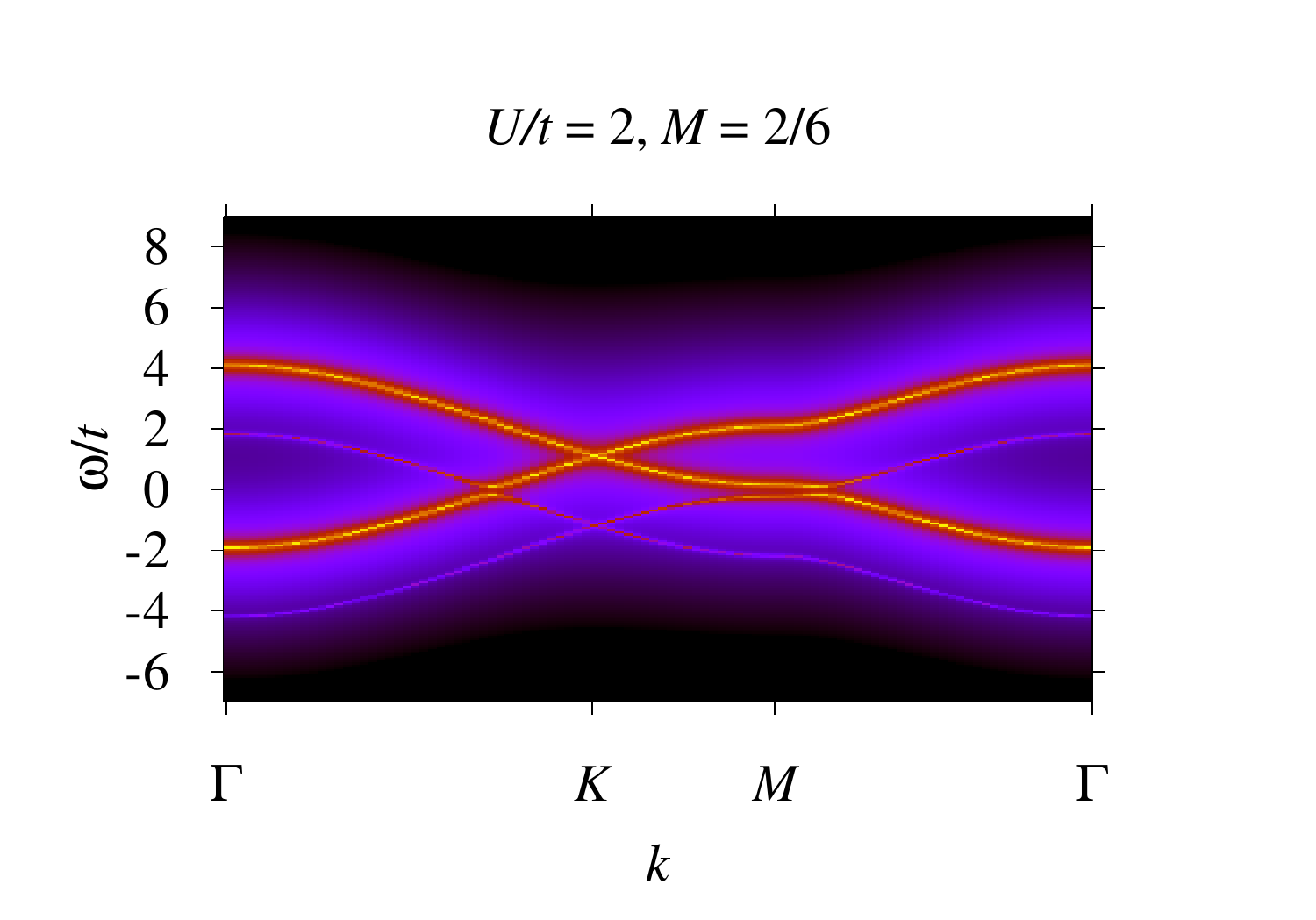}}
   \subfigure[$$]{\includegraphics[width=0.49\textwidth]{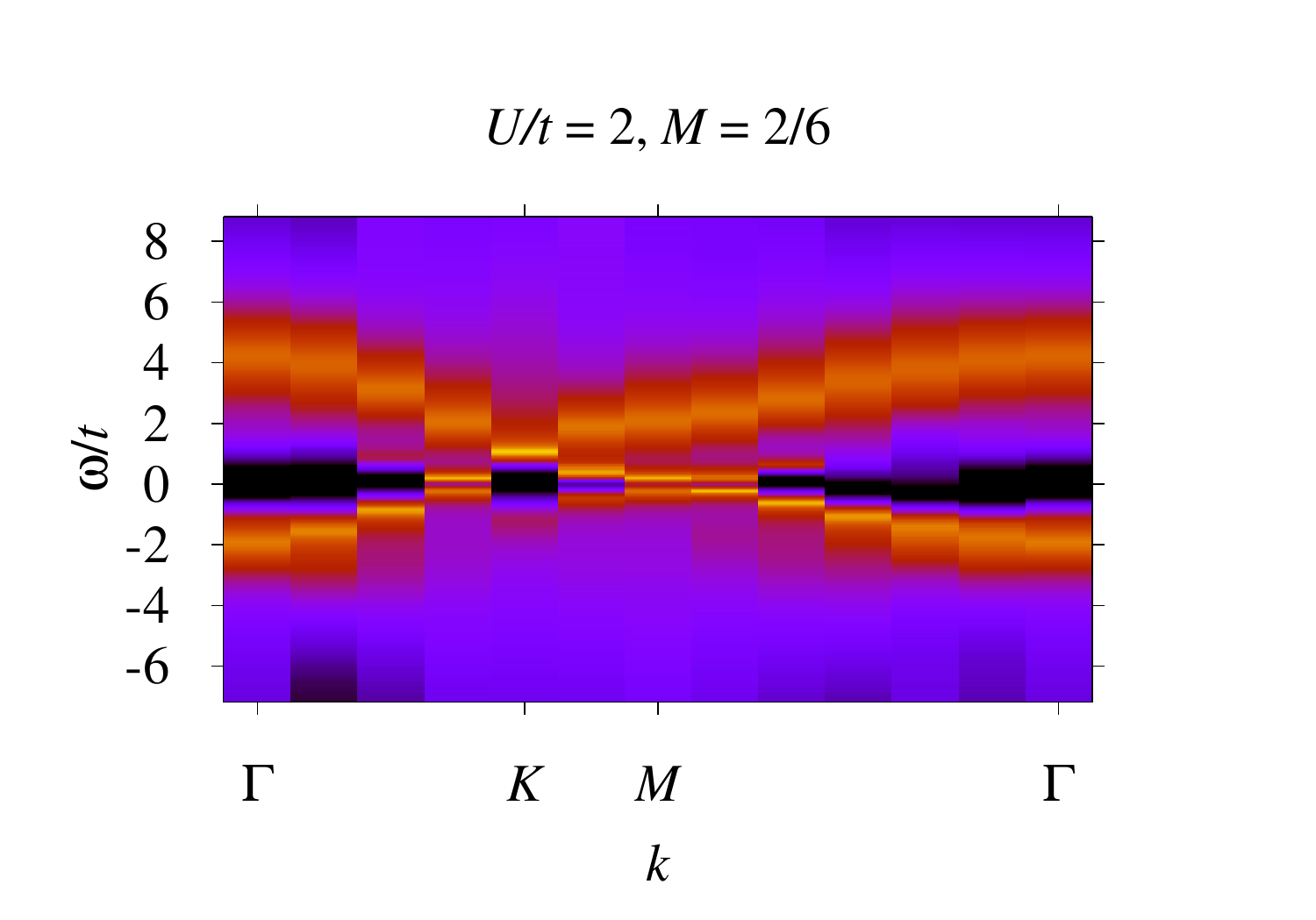}}
   \subfigure[$$]{\includegraphics[width=0.49\textwidth]{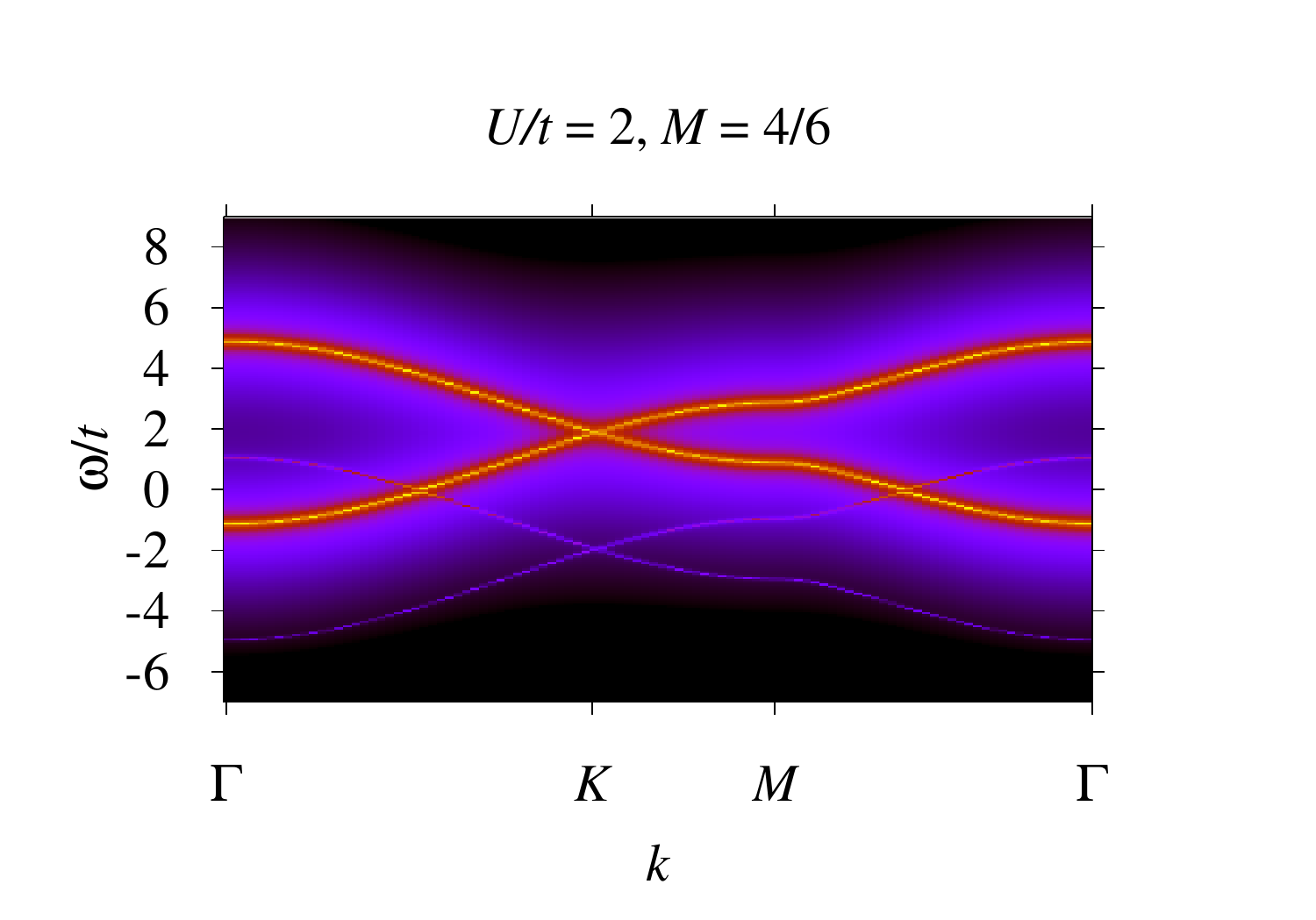}}
   \subfigure[$$]{\includegraphics[width=0.49\textwidth]{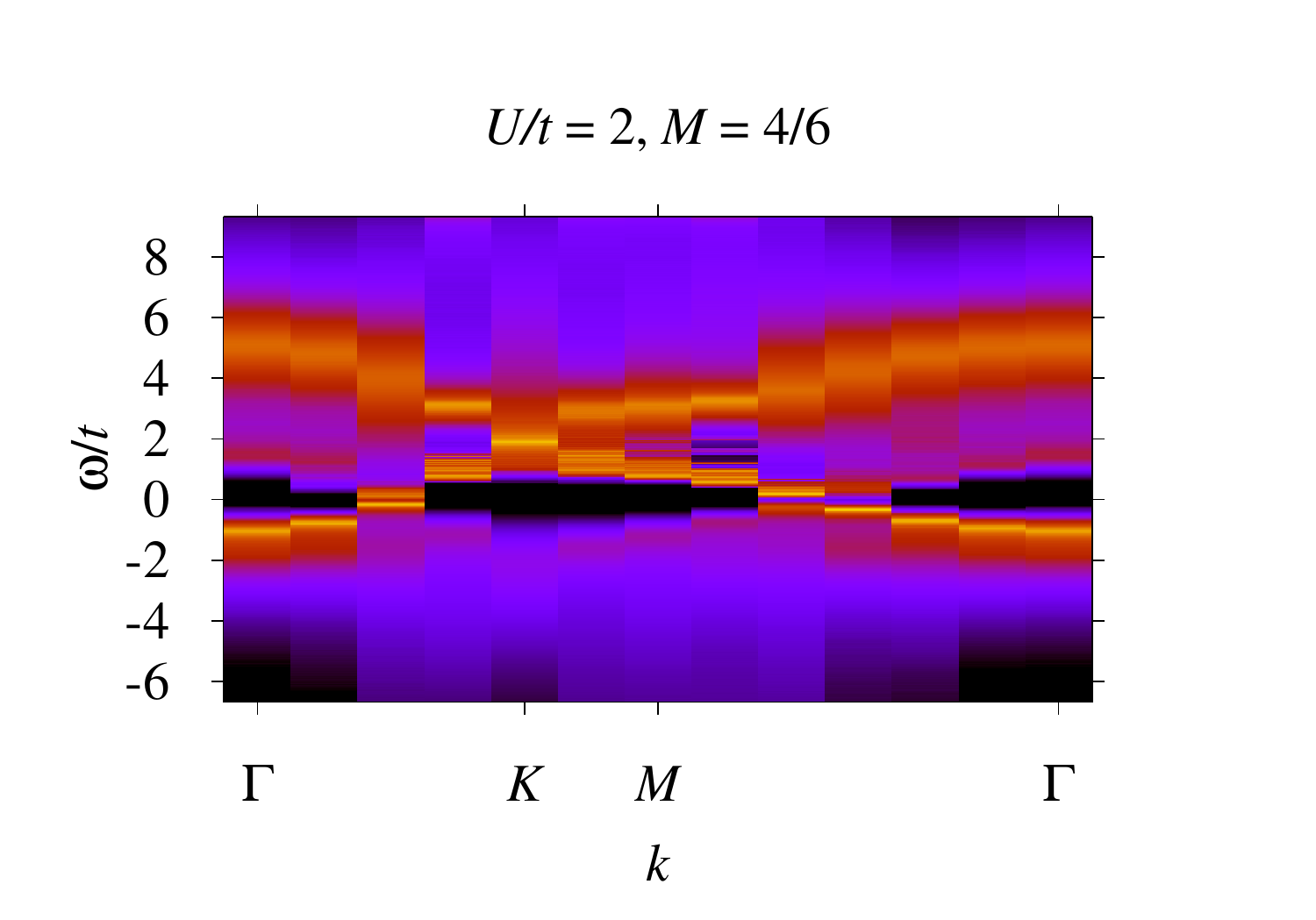}}
   \caption{Single-particle spectral function
${A^{\uparrow}(\mathbf{k},\omega)}$ at $U/t=2$, based on the
mean-field (left) and the QMC (right) calculations, respectively. The
magnetization $M$ takes the
values of $0$, $2/6$, and $4/6$ (from top to bottom). For the intermediate
value of the magnetization, one can clearly see the opening of a quasiparticle gap and the avoided
level crossing as $m_{\perp}$
acquires a finite value. In general, the magnitude of the gap tracks
$m_\perp$ [Fig.~\ref{fig5}(a)]. The QMC
spectral functions were obtained via analytical continuation of the
Green's functions with the
stochastic maximum entropy method. For the QMC calculations, the lattice
size was  set to $12\times
12$ unit cells. The legends at the top indicate the false color values.
   \label{fig6}}  
\end{figure*}

Our mean-field results are plotted in Figs.~\ref{fig4}, \ref{fig6}(a), \ref{fig6}(c),  and \ref{fig6}(e).  At zero magnetic field, we observed
as a function of  $U/t$ the expected transition from the semimetallic state (${\bf m} = 0 $)  at ${U/t < U_c/t \approx 3.3}$ to the antiferromagnetic  
Slater insulator characterized by ${|m_{\perp}| > 0}$.  
The semimetallic state at $B=0$ is characterized by the spin degenerate dispersion relation, $ \lambda_{n,\sigma} ({\bf k})$,  
as shown in Fig.~\ref{fig6}(a). Ramping up the magnetic field  lifts this degeneracy  thereby producing  nested  Fermi sheets
of opposite spin indices.  Hence, and irrespective  of the  magnitude  of $U < U_c$,   energy can be gained by ordering the
spins  in a canted antiferromagnet.  The energy gain corresponds to the gap which in the weak coupling limit, and by virtue of
Eq.~(\ref{gapeqn1}), takes the  form
\begin{equation}
\Delta_{\perp}\cong W e^{-3/2 U \rho(B_{\mathrm{eff}})}.
\end{equation}
The dispersion relation of this canted antiferromagnetic  state is plotted in Fig.~\ref{fig6}(c). To compare at best with  the QMC simulations we consider the quantity
\begin{equation}
\label{Aom.eq}
     A^{\uparrow} ({\pmb k}, \omega ) =  -\frac{1}{\pi}  {\rm Im } \left[
            G^{\uparrow}_{aa}({\pmb k}, \omega )   +   G^{\uparrow}_{bb}({\pmb k}, \omega ) \right]  
\end{equation}
with a finite broadening. As apparent,  the features with  dominant weight follow  the dispersion
relation $ \lambda_{1,\uparrow} ({\bf k})$ and $ \lambda_{2,\uparrow} ({\bf k}) $ and a gap at the
Fermi level is apparent.  Due to the  transverse staggered moment, mixing between the up and the down dispersion
relations occurs  thereby generating  shadow features following the dispersion relations of
$ \lambda_{1,\downarrow} ({\bf k})$ and $ \lambda_{2,\downarrow} ({\bf k}) $. The  intensity of   the shadow
features  tracks  $m_\perp$.   As apparent  from Fig.~\ref{fig4}  the  growth of $m_{\perp}$  as a function of the
magnetic field is countered by the polarization of the spins along the magnetic field. It is interesting to note that
irrespective of  $U/t$ the maximal value of $m_\perp$  and hence of the  magnetic field induced gap  is at
$B = 1$  corresponding to an energy scale matching  the position of the Van Hove singularity in the
non-interacting density of states. At this point a maximal amount of energy can be gained by the opening of the gap.

Particle-hole symmetry can be exploited to map the repulsive Hubbard model
onto an attractive Hubbard model where ${U < 0}$
tunes the transition from a semimetal to an $s$-wave superconductor. \cite{Scalettar89}
The external magnetic field driving the semimetal to a canted
antiferromagnet in the positive-$U$ model translates to
doping the negative-$U$ model which triggers a transition to a
uniform superfluid phase with $s$-wave pairing, accordingly.
\cite{Zhao06} In the case of $B=0$ the self-consistent equation for the staggered magnetization, Eq.~(\ref{gapeqn1}), is identical to the BCS gap equation for the attractive-$U$ case at half filling.
\section{\label{sec:section3}Projector quantum Monte-Carlo method}
To confirm   our mean-field results, we have carried out projector auxiliary field QMC calculations.
This PQMC algorithm is  based on the equation
\begin{equation}\label{projobs}
\frac{\langle\Psi_{0}|A|\Psi_{0}\rangle}{\langle\Psi_{0}|\Psi_{0}\rangle}
=\underset{\theta\rightarrow\infty}{lim}\frac{\langle\Psi_{T}|e^{-\theta H}Ae^{-\theta H}|\Psi_{T}\rangle}{\langle\Psi_{T}|e^{-2\theta H}|\Psi_{T}\rangle}\;.
\end{equation}
The trial wave function $| \psi_{T} \rangle $ has to be non-orthogonal to the ground state wave function,
$\langle \psi_{T}|\psi_{0}\rangle\neq 0$ and the ground state is assumed to be non-degenerate.    For the details
of the formulation of this approach, we refer the reader to Ref. \cite{Assaad08b}.
In this canonical approach, we fix the  magnetization
\begin{equation}
     M = \frac{ N_{\uparrow}  -  N_{\downarrow} }{N_{\uparrow}  +  N_{\downarrow} }
\end{equation}
rather than the magnetic field (just as a magnetic field would induce a magnetization). The magnetization $M$ in the QMC algorithm corresponds to the mean-field magnetization $m_{\parallel}$. $N_{\sigma}$ corresponds to the total number of electrons in the spin sector $ \sigma$.  Furthermore, due to the particle-hole symmetry, which locks  in the signs of the fermionic
determinants in both spin sectors one  can avoid the so-called negative sign problem irrespective
of the choice of the magnetization.
In practice,  for each finite system, we choose a value of the projection parameter
$\theta$ large enough so as  to guarantee  convergence within  statistical uncertainty.  

\begin{figure}[htbp]
   \centering
   \subfigure[$$]{\includegraphics[width=0.5\textwidth, angle=-90]{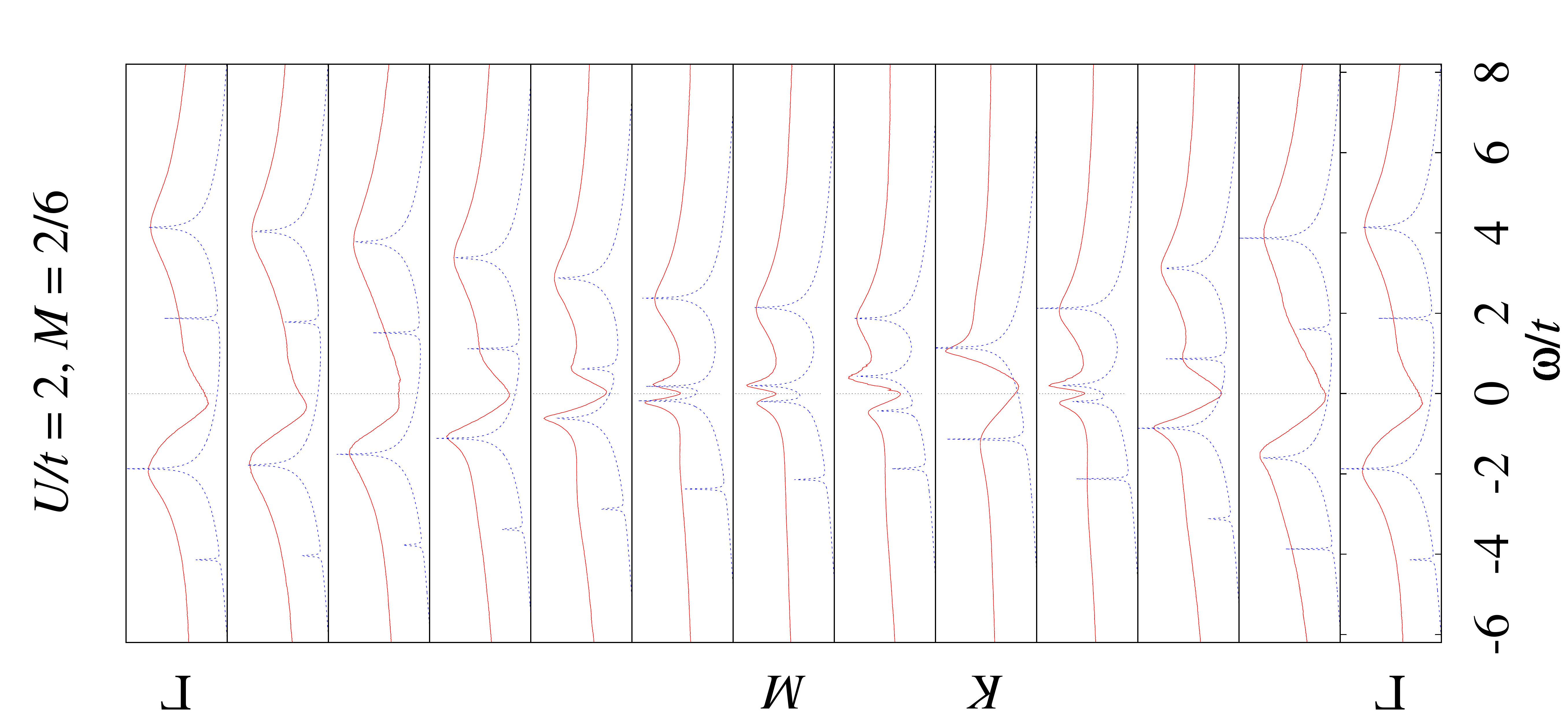}}
   \subfigure[$$]{\includegraphics[width=0.5\textwidth, angle=-90]{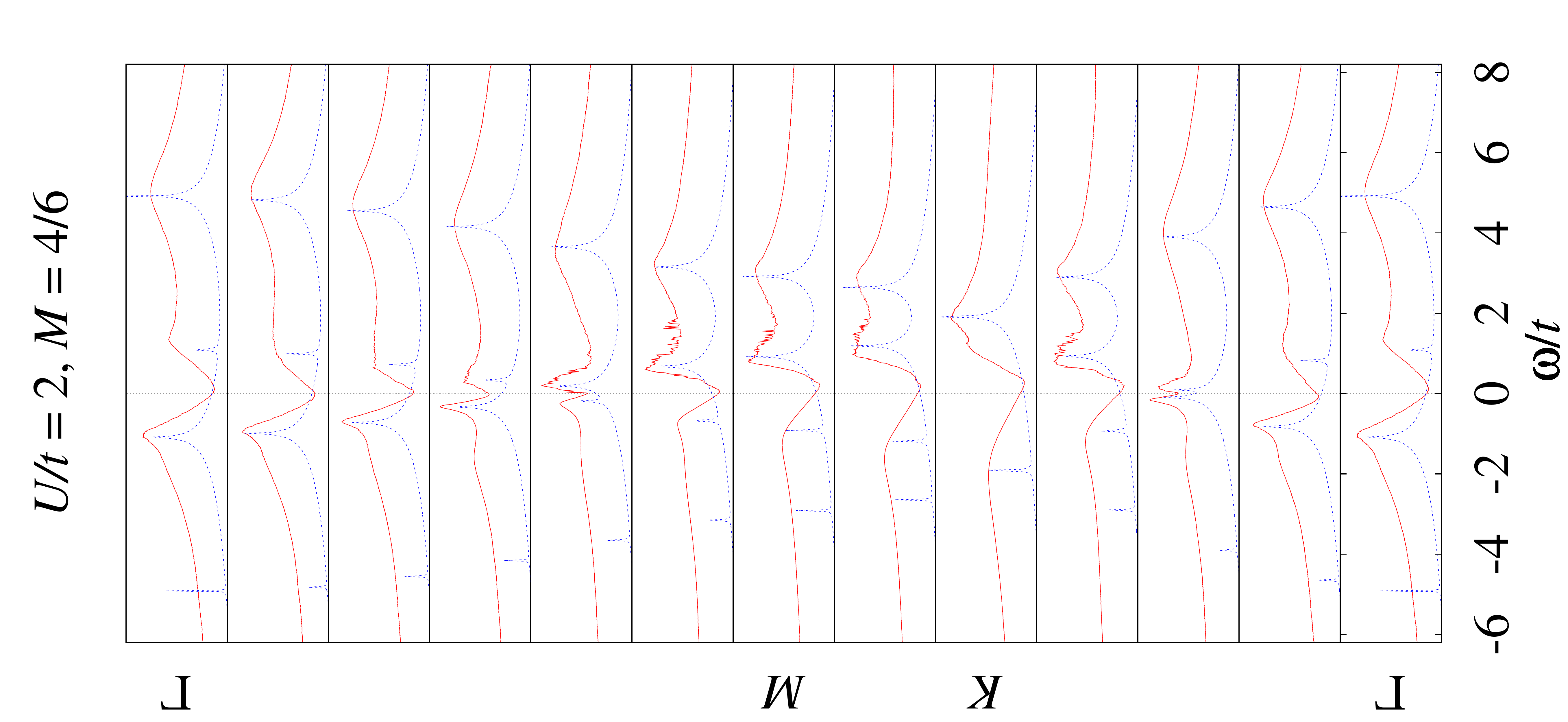}}
   \caption{The plot directly compares the QMC (red) spectral intensity profiles $A^{\uparrow}(\mathbf{k},\omega)$ against the mean-field solution (blue, dotted). Same parameters as in Fig.~\ref{fig6}.
   \label{fig7}}
\end{figure}
To detect transverse staggered magnetic order under an applied magnetic field, we have computed the
spin-spin correlation functions
\begin{equation}
S^{+-}( \mathbf{q} )
=\frac{1}{N}\sum_{\mathbf{i},\mathbf{j}} e^{-\i \mathbf{q}(\mathbf{i}-\mathbf{j})}
\langle S^{+}_{a}(\mathbf{i}) S^{-}_{a} (\mathbf{j})    
     -  S^{+}_{a}(\mathbf{i}) S^{-}_{b} (\mathbf{j})
     \rangle\;.
\end{equation}
If long-range staggered magnetic order perpendicular to  the applied field direction   is present,
then  
\begin{equation}
     \sqrt{\lim_{N \rightarrow \infty}  \frac{S^{+-}(\mathbf{Q})}{N} } =  m^{\mathrm{QMC}}_\perp  
\end{equation}
acquires a finite value.
We have computed this quantity on $6\times 6$, $9\times 9$, and $12\times 12$ lattices, and our
results are plotted  in  Fig.~\ref{fig5} both for $ U < U_c $ and $ U > U_c $.  
At $ U/t = 2 < U_c/t$  and zero  magnetization, $M =0$,  our results are consistent with  $ m^{\mathrm{QMC}}_\perp  =  0$  
whereas, at $ M = 2/6 $,  $ m^{\mathrm{QMC}}_\perp $  takes a finite value. Although we cannot reproduce
the  essential singularity of the  mean-field  calculation at $U < U_c$, the overall form of the
transverse staggered     magnetization  compares favorably with the  mean-field results
[see Figs.~\ref{fig5}(a) and ~\ref{fig5}(b)] both  at  $ U < U_c $ and $ U > U_c $.

Within the PQMC, the zero-temperature  single-particle Green's function along the
imaginary time axis  can be computed efficiently  with methods introduced in Ref.~\cite{Feldbacher01}.
From this quantity, we can obtain the spectral function of Eq.~(\ref{Aom.eq}) with the use of
a stochastic  formulation of the maximum entropy method. \cite{Beach04a,Sandvik98}  The so
obtained results for $A^{\uparrow}({\bf k}, \omega) $ are plotted in Figs.~\ref{fig6}(b), ~\ref{fig6}(d), and ~\ref{fig6}(f).
As apparent the features in the QMC calculation which are associated with substantial
spectral weight are well reproduced by the mean-field calculation.  
The particle-hole transformation,
$ {\hat a}^{\dagger}_{ {\bf i}, \sigma } \rightarrow   {\hat a}_{ {\bf i}, -\sigma }  $ and
$ {\hat b}^{\dagger}_{ {\bf i}, \sigma } \rightarrow   -{\hat b}_{ {\bf i}, -\sigma }  $,
leads to the relation
\begin{equation}
     A^{\uparrow}( {\bf k} , \omega ) = A^{\downarrow}( {\bf k} , -\omega ).
\end{equation}
At finite magnetic fields or equivalently at finite magnetizations, the staggered transverse
order leads to a gapless Goldstone mode of which the quasiparticle can spin-flip scatter.
As a consequence, and as already observed in the mean-field calculation,  
the features of the  down spectral function should be visible in  
$ A^{\uparrow} ({\bf k}, \omega) $.   Upon inspection of Fig.~\ref{fig7} one will observe that
for each  dominant low-energy peak at $ \omega({\bf k} )$  in  $ A^{\uparrow} ({\bf k}, \omega) $
a shadow feature at $-\omega({\bf k} )$ is present.
\begin{figure}[htbp]
   \centering
   \includegraphics[width=0.45\textwidth]{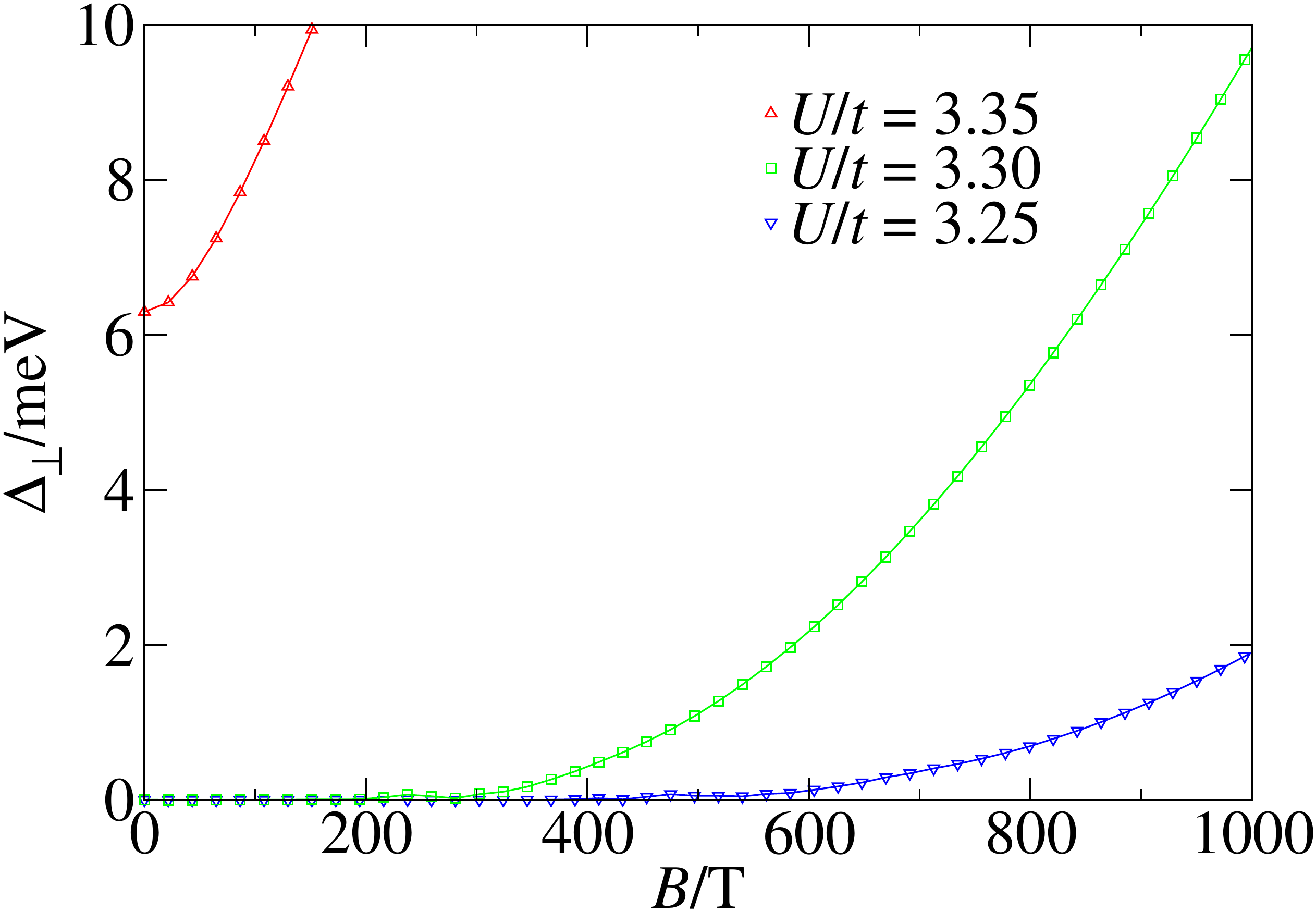}
   \caption{Charge gap $\Delta_{\perp}$ as a function of the applied magnetic field
   for $t=2.5$eV, obtained by numerically solving the gap equations, (\ref{gapeqn}).
   \label{fig8}}
\end{figure}


\section{ \label{sec:section4} Summary and Conclusions }

In conclusion, we have carried out  mean-field  calculations and projective quantum Monte Carlo  simulations
for the Hubbard model on the honeycomb lattice in a  magnetic field oriented parallel to the lattice plane. 
With this setup, only the Zeeman  spin coupling is present. Our results show the inherent instability of the
semimetallic state to a canted antiferromagnet upon application of the magnetic field. Ramping up the
magnetic field generates  nested  up and down Fermi surfaces with finite density of states.  As a consequence,
the onset of canted antiferromagnetic order opens a charge gap and provides an energy gain irrespective of
the magnitude of the Hubbard repulsion.

Experimentally, such a transition could be observed by magneto resistance measurements. The transition to
the canted  antiferromagnet breaks a U($1$) symmetry and hence occurs at finite temperatures
$T_{\mathrm{KT}}$ in terms of a Kosterlitz-Thouless transition.  Below $T_{\mathrm{KT}}$ the power-law decay
of the transverse spin-spin correlation function should suffice to produce a  visible  pseudo-gap in the
charge sector and hence an increase in the resistivity as a function of decreasing temperature. The primary
issue to observe the transition is the magnitude of the required  magnetic field so as to obtain a visible
gap.  With  $t\approx 2.5-3.0$eV and $U\approx 10-16$eV, \cite{Parr50Baer86} one can readily see that very large magnetic
fields will  be required to  obtain charge gaps  in the meV region. In particular, in Fig.~\ref{fig8} we plot the
charge gap in meV as a function of $B$ in tesla for values of the Coulomb repulsion close to $U_{c}$. The
g factor has been set to the (approximate) free-electron value, $g=2$. As apparent, depending on $U$, values on the order
of $ B \propto  10^{2}-10^{3}$T  are required to obtain an acceptable gap. Clearly those numbers imply that 
the only feasible manner to observe this effect would be to grow graphene directly on a magnetic substrate. 

With those numbers in mind, we can now consider magneto resistance experiments  carried out  in layered highly oriented
pyrolitic graphite. \cite{Kempa03}  For  magnetic fields perpendicular to the plane  and at low  temperatures,   an
{\it insulating } state   as characterized by $\frac{d\rho}{dT}$ is observed at ${B \propto 0.1}$T.
For fields parallel to the plane, intensities of roughly $B \propto 10$T are required to observe the insulating behavior in the
magneto resistance.  Given this data, it is clear that the  dominant effect  of the magnetic field stems from the
orbital coupling rather than from the  Zeeman spin coupling. \cite{Khveshchenko01}   
The authors of Ref. \cite{Kempa03}  account for the parallel field data  by mentioning deviations from perfect
alignment between the graphene plane and the magnetic field.  Given our estimate of the required magnetic field
to achieve a charge gap  for the parallel field configuration, we can only confirm this point of view.

\begin{acknowledgements}
We would like to thank B. Trauzettel for helpful comments and discussions. The simulations were carried
out on the HLRB2 at the LRZ, Munich. We thank this institution for generous allocation of CPU time.
Financial support by the DFG under Grant No. AS120/4-2 is acknowledged.
\end{acknowledgements}

\bibliographystyle{./apsrev}

\end{document}